\documentclass[a4paper,11pt]{article}
\pdfoutput=1 

\usepackage{jheppub} 
\usepackage[T1]{fontenc} 
\usepackage{graphicx}
\usepackage{bm,amsmath,amssymb}
\usepackage[mathscr]{eucal}

\long\def\comment#1{ }
\newcommand{\eqn}[1]{Eq.~(\ref{#1})}
\newcommand{\beq}{\begin{equation}}
\newcommand{\eeq}{\end{equation}}
\newcommand{\nn}{\nonumber\\}

\newcommand{\rmd}{{\rm d}}

\newcommand{\rmi}{{\rm i}}

\newcommand{\del}{\partial}

\newcommand{\order}[1]{\mcal{O}{(#1)}}
\newcommand{\mcal}{\mathcal}
\newcommand{\st}{t}
\newcommand{\sh}{h}

\title{A semi-holographic model for heavy-ion collisions}

\author[a]{Edmond Iancu}
\author[a,b,c]{and Ayan Mukhopadhyay}

\affiliation[a]{Institut de Physique Th\'{e}orique, CEA Saclay, F-91191 Gif-sur-Yvette, France}
\affiliation[b]{Centre de Physique Th\'{e}orique, Ecole Polytechnique, CNRS, 91128 Palaiseau Cedex, France
}
\affiliation[c]{Crete Center for Theoretical Physics (CCTP) and Crete Center for Quantum Complexity and Nanotechnology (CCQCN), University of Crete, P.O. Box 2208, 71003, Heraklion, Greece \footnote{Present address of A.M. since October 1, 2014}}

\emailAdd{edmond.iancu@cea.fr}
\emailAdd{ayan@physics.uoc.gr}

\abstract{We develop a semi-holographic model for the out-of-equilibrium dynamics during the partonic
stages of an ultrarelativistic heavy-ion collision. The model combines a weakly-coupled hard sector, 
involving gluon modes with energy and momenta of the order of the saturation momentum and
relatively large occupation numbers, with a strongly-coupled soft sector, which physically represents the
soft gluons radiated by the hard partons. The hard sector is described by perturbative QCD,  more
precisely, by its semi-classical approximation (the classical Yang-Mills equations) which becomes
appropriate when the occupation numbers are large. The soft sector is described by a
marginally deformed conformal field theory, which in turn admits a holographic description in terms of
classical Einstein's equations in $AdS_5$ with a minimally coupled massless `dilaton'.
The model involve two free parameters which characterize
the gauge-invariant couplings between the hard and soft sectors. Via these couplings, the hard
modes provide dynamical sources for the gravitational equations at the boundary of $AdS_5$ and
feel the feedback of the latter as additional soft sources in the classical Yang-Mills equations. Importantly,
the initial conditions for this coupled dynamics are fully determined by the hard sector alone, i.e. by
perturbative QCD, and are conveniently given by the color glass condensate (CGC) effective theory. 
We also develop a new semi-holographic picture of jets in the QGP by attaching a non-Abelian charge to the endpoint of the trailing string in $AdS_5$ representing a heavy quark. This leads to modified Nambu-Goto equations 
for the string which govern the (collisional and radiative) energy loss by the heavy quark towards both hard and soft modes.
}

\begin{document} 
\maketitle
\flushbottom

\section{Introduction}
The experimental heavy ion programs at RHIC and the LHC have brought compelling evidence that the
dense partonic form of matter known as the `quark--gluon plasma' (QGP) created in the intermediate stages of a
ultrarelativistic nucleus--nucleus collision behaves in many respects like a `perfect fluid' ---
a many--body system with strong interactions. Perhaps the most suggestive data in that sense
are those referring to collective phenomena like elliptic flow, which, via their theoretical description
in terms of hydrodynamics, point out towards a short thermalization time and a remarkably small
value for the ratio between viscosity and entropy density \cite{Teaney:2009qa,Ollitrault:2012cm} (and Refs. therein),
as expected at strong coupling.
This behavior can be attributed to either large gluon occupation numbers leading to strong non--linear 
phenomena even when the coupling is weak, 
or to a genuinely non--perturbative regime, associated with relatively soft gluons, for which
the QCD running coupling is indeed quite strong. Most likely, both features combine to yield the 
observed `perfect liquid' behavior, and their interplay is expected to vary with time. 

The gluonic
matter liberated in the early stages of the collision is characterized by large occupation numbers, 
of order $1/\alpha_s$, and by a relatively weak value of the QCD running coupling
$\alpha_s=g^2/4\pi$ --- precisely because the density is so high.
Via the gluon non--linear dynamics, this high density introduces a semi--hard scale, the `saturation 
momentum' $Q_s$, which is the typical momentum of the liberated gluons and also the relevant scale 
for evaluating the running coupling: $\alpha_s(Q_s)\ll 1$. But although the coupling is weak,
the mutual interactions of this dense gluonic matter 
-- also called the `color glass condensate' (CGC) \cite{Iancu:2002xk,Gelis:2010nm}
are effectively of order one: the factors of $\alpha_s$ from the interaction
vertices are compensated by the large occupation number $\sim 1/\alpha_s$ of the interacting modes. 
With increasing time, this matter rapidly expands and thus gets diluted,
while at the same time abundantly emitting relatively soft gluons, with energies and momenta
well below $Q_s$. These soft gluons have occupation numbers of order one or less, but a
relatively stronger coupling, perhaps as strong as $\alpha_s\sim 1$, 
and their mutual interactions are expected to represent
the driving force towards thermalization. In particular,  the semi--hard partons eventually thermalize
because they lose energy towards the thermal bath made by the soft gluons. This
out--of--equilibrium partonic matter, sometimes referred to as `glasma' \cite{Lappi:2006fp},
can also include {\em very} hard partons, with energies and/or masses
much larger than $Q_s$, which are created via binary collisions between the constituents of
the incoming nuclei. Such `hard probes' couple to both the semi--hard and the soft constituents
of the surrounding medium, possibly in different ways, and thus can lose energy, or get deflected.

It would be highly desirable to have a unique theoretical description for all the aspects
of this complicated non--equilibrium dynamics, but this seems difficult to achieve from 
first principles. So far, most studies have used a
one--sided approach to the problem, which privileges either a weak--coupling scenario, 
or a strong--coupling one. 

In QCD at weak coupling, one has been able to deal with the highly--occupied
gluon modes via semiclassical techniques, like numerical solutions to the classical Yang--Mills 
equations supplemented with fluctuations in the initial conditions 
(and possibly coupled to non--Abelian Vlasov equations for the harder modes)
\cite{Romatschke:2006nk,Dumitru:2006pz,Attems:2012js,Berges:2012cj,Berges:2013eia,Gelis:2013rba}. 
An alternative strategy, which becomes appropriate when the occupation numbers drop down to values
much smaller than $1/\alpha_s$, involves analytic and numerical studies of appropriate transport equations, 
which include gluon radiation \cite{Baier:2000sb,Arnold:2002ja,Arnold:2002zm,Kurkela:2011ti,Kurkela:2011ub,Kurkela:2014tea}.
Whereas the conventional wisdom, based on the experience with kinetic theory, 
is that the thermalization time at weak coupling should be parametrically large 
\cite{Baier:2000sb,Kurkela:2011ti,Kurkela:2011ub} --- due to the smallness
of the cross--sections for partonic interactions (and despite the presence of plasma 
instabilities \cite{Attems:2012js}) ---, more recent numerical studies 
\cite{Gelis:2013rba,Kurkela:2014tea} suggest that rapid thermalization
is not necessarily in contradiction with the weak coupling picture.
For instance, Ref.~\cite{Gelis:2013rba} observed a relatively fast isotropization of the 
pressure tensor due to the non--linear dynamics of strong classical Yang--Mills fields.

At strong coupling, on the other hand, it is customary to rely on the AdS/CFT correspondence, where
thermalization in the conformal ${\mathcal N}=4$ supersymmetric Yang--Mills (SYM) theory is described
as the formation of a black--brane horizon in the radial direction of the dual $AdS_5$ space--time.
This process has been studied via both analytic \cite{Janik:2005zt,Lin:2006rf,Kovchegov:2007pq,Gubser:2008pc,Albacete:2008vs,Lin:2008rw,Beuf:2009cx} and numerical 
\cite{Chesler:2008hg,Chesler:2009cy,Chesler:2010bi,Wu:2011yd,Heller:2012km,Wu:2012rib,Casalderrey-Solana:2013aba,
vanderSchee:2013pia,Chesler:2013lia,vanderSchee:2014qwa} methods, but in a purely holographic
set--up, that is, starting with initial conditions which describe some non--equilibrium state at strong coupling, whose relevance for the heavy--ion problem is difficult to assert. Perhaps the closest to the heavy--ion 
scenario is the scattering between two gravitational shock--waves, but even in that case the 
correspondence to the actual physical situation (the scattering between two dense systems 
of gluons which are weakly--coupled) is far from being clear, and not truly under control.

The need for combining insights and also techniques from both weak and strong--coupling approaches,
with the aim of developing a unified approach, has since long been recognized (see e.g. the discussion
in \cite{Mueller:2008zt}). However, so far it has only been implemented  \cite{Casalderrey-Solana:2014bpa} 
in the form of a hybrid model for  `jet quenching' --- the energy loss by an energetic parton propagating
through a strongly--coupled plasma. In Ref.~ \cite{Casalderrey-Solana:2014bpa},  
the fragmentation of the incoming parton has
been described within perturbative QCD, while the interactions between the ensuing parton cascade
and the surrounding medium have been modeled by using results from the AdS/CFT 
correspondence, that is, by assuming the plasma to be exclusively made with strongly interacting modes.

In this paper, we would like to propose a more general strategy for simultaneously describing
weak--coupling aspects and strongly--coupling ones, that we shall refer to as 
{\em semi--holography}. This aims at keeping both the (semi)hard and soft
degrees of freedom in the medium, together with their various types of interactions: weak interactions
among the hard partons, as described by perturbative QCD (including the resummations needed
to deal with large gluon occupation numbers), strong interactions among the soft modes, as
encoded into a holographic model to be shortly explained, and `hard--soft' interactions which 
couple the two types of degrees of freedom in a minimal way.  As we shall see, our model involves
only a small number of free parameters (two parameters related to hard-soft couplings, plus an additional parameter associated with the  hard probes), and therefore can be falsified. 

Remarkably, our description can be applied to all the 
stages of an ultrarelativistic nucleus--nucleus collision, from the initial conditions (which are
exactly the same as in the standard CGC approach in pQCD) until hadronization. 
Albeit it reduces to classical equations of motion, the dynamics generated by our model is quite complex. 
Due to the specificity of the holographic initial and boundary value problem, the coupled equations of motion
for the hard and soft sectors should be solved in a self-consistent manner, via numerical iterations. 
Our description naturally ends when the effective temperature drops below the confinement scale, meaning at the onset of hadronization.

The holographic part of our model is a minimal version
of classical gravity in $AdS_5$, which includes the graviton and a massless scalar field
(the `dilaton'). This gravitational theory in $AdS_5$ may be thought of 
as dual to an `infrared conformal field theory' (IR-CFT), which is
strongly coupled and provide a model for the soft, non--perturbative degrees of freedom 
in the QGP phase of QCD. This is completed by the `hard--soft' interactions, which represent marginal 
deformations of the strongly--coupled IR-CFT, and in the dual description, provide boundary
sources for the graviton and dilaton fields in the bulk.

To summarize, the semi--holographic model contains classical Yang-Mills fields representing the hard sector,
(`hard gluons', or `glasma fields') coupled to a few operators of a strongly coupled IR-CFT, whose spacetime 
evolution is self--consistently given by the equations of classical gravity in $AdS_5$ with dynamical sources 
on the boundary. The couplings between the glasma fields and the IR-CFT operators involve gauge--invariant, 
local operators of lowest possible scaling dimensions selected according to their quantum numbers. We need only two such operators both in the hard and soft sectors -- namely 
the energy momentum tensor and the `glueball' operator (the Yang--Mills Lagrangian density). 
We will derive equations for the coupled evolution of the hard and soft sectors, 
with initial conditions entirely fixed by the CGC description of the hard modes, and devise
a numerical strategy for solving these coupled equations.

Additionally we will also provide a semi--holographic model for the energy loss by a heavy quark.
This quark simultaneously couple to the perturbative, glasma, fields --- via its non--Abelian, `color', charge,
whose dynamics is modeled by the Wong equation --- and to the soft, strongly--coupled, sector ---
via its virtual cloud of soft quanta, as captured by a trailing string in $AdS_5$ attached to 
the non--Abelian charge at the boundary. The Nambu--Goto dynamics of this string in the evolving bulk
geometry leads an interesting generalization of the non--Abelian Lorentz force acting on the
heavy quark, which encompasses both the respective force due to the classical glasma fields, and the 
flow of energy and momentum towards the soft degrees of freedom. Using the method in 
\cite{Casalderrey-Solana:2014bpa},  one can extend our model to include the perturbative 
fragmentation of the heavy quark into quarks and gluons,
and thus provide a semi--holographic picture for the evolution of a jet as a whole.

A semi--holographic approach has been previously applied to condensed matter systems, with a motivation
similar to ours --- namely, to build an effective theory for a large class of emergent non--Fermi liquids close to quantum criticality, integrating both weak and strong--coupling aspects 
\cite{Faulkner:2010tq, Mukhopadhyay:2013dqa}. In particular a generalization of Landau's theory for these non--Fermi liquids has been proposed \cite{Mukhopadhyay:2013dqa}, where the entire low energy phenomenology is described in terms of generalized Landau parameters, which may vary from one system to another. However, there are
some noticeable differences between such previous approaches and our present one: in the context of condensed
matter, the soft degrees of freedom are parametrically denser (by a factor $N^2$) than the hard degrees of freedom in the large--$N$ limit\footnote{The parameter $N^2$, with $N$ playing the role of the rank of the non--Abelian gauge group in the emerging IR-CFT theory,  is related to the inverse of the gravitational constant in the holographic description.}. Accordingly, in that case, one can ignore the back-reaction of the hard degrees of freedom on the soft sector, which greatly simplifies  the analysis of the latter.

By contrast, in our semi--holographic model for the QGP, the densities of both hard and soft modes should scale in the same way with $N$, since they both count the number of gluon--like degrees of freedom\footnote{In the present, heavy ion, context, this parameter $N$ should be related to the rank $N_c$ of the `color' gauge group in QCD; see the discussion in the next section.} (with however have different energies and intrinsic gauge couplings). This feature not only complicates the dynamics in our model,
but also introduces some ambiguity in its formal structure: unlike in the corresponding discussion of non--Fermi liquids \cite{Mukhopadhyay:2013dqa}, here it is not possible anymore to distinguish between relevant and irrelevant operators in the large $N$ limit. In view of that, we have opted for a `minimalist' model involving the smallest number
of hard--soft couplings, which is two. One additional argument behind this choice, 
is the fact that couplings involving ``multi-trace IR-CFT operators'' should be dynamically suppressed by 
powers of the hard saturation scale $Q_s$. In Section 3, we will also present a renormalization group argument which gives more support to our model.

The plan of the paper is as follows. In Section 2, we will present our semi--holographic model, meaning the full set of coupled equations for the hard and soft degrees of freedom. We will then discuss the initial conditions 
(for both hard and soft sectors) and explain that they can be unambiguously inferred from the CGC description of the colliding nuclei. We will also describe the iterative numerical strategy for solving the equations. In Section 3, we will present a renormalization group flow argument outlining the set of assumptions under which our model will turn out to be an effective theory for the QGP phase. In Section 4, we will present our semi-holographic model for jets. In the concluding Section, we will mention some key directions which can be readily pursued and some further possible simplifications of our model that may be easier to implement.

\section{The model}
In the early stages of heavy ion collisions in QCD, 
the hard scale is set by the saturation scale $Q_s$ of the incoming nuclei. 
This is so because the partons (mostly gluons) released just after the collision,
at a proper time $\tau\sim 1/Q_s$,
have both transverse and longitudinal momenta of the order of $Q_s$. The QCD
coupling at this scale is generally assumed to be reasonably weak, $\alpha(Q_s)\ll 1$, for
perturbation theory to apply\footnote{In practice,  $\alpha(Q_s)\simeq 0.3$, so this condition
is at most marginally satisfied. However, experience with perturbative QCD indicates that
weak coupling techniques, when supplemented with appropriate resummations, retain a good
predictive power for such a value $\alpha_s\sim 0.3$.}.  At times 
$\tau\lesssim 1/Q_s$,  the gluon modes have large occupation numbers,  of order 
$1/\alpha(Q_s)$, so the non--linear effects are strong even if the coupling is weak. 
This non--linear dynamics is well accounted for
by the classical Yang-Mills equations supplemented with a renormalization group
equation for the distribution of their `sources' (the color charge densities in the incoming nuclei).
This picture lies at the basis of a semi--classical effective theory for the initial and
early stages of a heavy ion collision, the `color glass condensate' \cite{Iancu:2002xk,Gelis:2010nm},
which is derived from pQCD.

However, the classical Yang-Mills equations cannot give an accurate description of the 
dynamics at later stages, for at least two reasons: on one hand, the partonic
system becomes more and more dilute, and hence under--occupied, 
due to rapid longitudinal expansion; on the
other hand, soft gluons are abundantly radiated and they are expected to interact rather
strongly with each other since the strength
of their mutual coupling is relatively large. These strong interactions should bring the soft modes
close to thermal equilibrium, albeit at a rather low temperature to start with, 
because of their comparatively low energy density. 
Subsequently, this soft thermal bath will continuously absorb energy from the reservoir made
by the hard particles and thus increase its temperature. At the same time, the hard partons will
lose energy towards the soft modes, until they reach equilibrium with the latter.

As explained in the Introduction, we here adopt the point of view that the soft gluon modes are 
strongly coupled among themselves and that their internal dynamics can be effectively 
described by an infrared conformal field theory (IR-CFT), which admits a 
holographic dual. More precisely, this
effective description applies to the intermediate--energy sector between the saturation
($Q_s$) and the confinement ($\Lambda$) scales, that we assume to be well separated
from each other: $Q_s\gg \Lambda$.

We shall require this emergent IR-CFT to be a proper conformal field theory, 
which exhibits scaling symmetry and Lorentz invariance, and for which one can take the limit
of a strong  't Hooft coupling. The Lorentz invariance is of course physically 
motivated --- this enables us e.g. to generate solutions which describe a boost invariant Bjorken flow
for arbitrary initial conditions. The assumption of scaling symmetry, on the other hand, and also the
strong--coupling limit, are rather crude idealizations, which are intended to simplify the problem.
They amount to neglecting the running of the QCD coupling within the intermediate energy range 
alluded to above, albeit this running was of course essential to justify the existence of a 
strong coupling in the first place.  Under these assumptions, we can consistently conjecture 
that the holographic dual of the IR-CFT is a supergravity theory 
living in an asymptotically $AdS_5$ space--time. 

In the above, we have implicitly assumed that the color gauge
group SU$(N_c)$ is the same for the IR-CFT as in QCD. However what we are really assuming here is that the $N_c$ of the gauge group of the IR-CFT is parametrically as large as 3 -- the $N_c$ of QCD, and of course $3$ is large enough to a good approximation to enable us to take a large $N_c$ limit. There are well known examples in supersymmetric QCD where a new gauge group emerges in the infrared \cite{Intriligator:1995au} which may or may not be strongly coupled. In our case, the holographic IR-CFT is strongly coupled and is assumed to emerge at intermediate scales in the quark-gluon plasma phase, where the effective coupling is expected to run very slowly. For consistency, we shall use the large $N_c$ limit also in the treatment of the `hard' 
gluon modes also, which are weakly coupled, and for which we shall keep the semi--classical QCD
description. However the  't Hooft coupling $\lambda \equiv g^2 N_c$ for the hard sector is small: $\lambda\ll 1$. 

We are now in a position to describe our semi--holographic model in more detail. This model
combines `hard' gluon modes, which are weakly coupled but highly populated, 
and are represented by classical Yang--Mills fields (also known in this context as
`glasma fields'), together with  `soft' and strongly 
coupled degrees of freedom, which are represented by gauge--invariant 
operators in the IR-CFT, and  whose dynamics can be studied with the help of the 
supergravity dual. The `internal' ('t Hooft) couplings in these two
sectors are treated as independent quantities, which are both fixed : in the hard
sector, the coupling is small, $\lambda\ll 1$ as mentioned above, whereas in the  IR-CFT, the respective
coupling is infinitely strong
and disappears from the problem in the classical gravity approximation.
These two types of degrees of freedom must be coupled with each
other via operators which become irrelevant in the `hard' limit $Q_s\to \infty$ -- as
in this limit the dynamics of the hard modes should become insensitive to the soft sector. Furthermore, the full hard and soft sectors must have a consistent large $N_c$ limit. 

Since the relevant operators of the IR-CFT are gauge invariant, the hard modes 
can couple to them only via gauge-invariant composite operators. A minimal model involves only two marginal operators from the IR-CFT, the energy-momentum tensor of the 
soft modes that we denote as $\mathcal{T}_{\mu\nu}$ and a scalar operator, denoted as
$\mathcal{H}$, which can be thought off as the glueball operator in the IR-CFT.
The classical gravity fields
dual to these operators are the graviton and a massless dilaton respectively. They obey 
the standard equations of classical gravity in the dual $AdS_5$ geometry, to be
presented below. The `partonic' operators which couple to these IR-CFT operators can
be inferred from symmetry arguments, together with the requirement that the hard--soft
couplings be suppressed by inverse powers of $Q_s$. 
As the expectation values of the soft sector operators will be vanishing at initial times (the soft sector can be produced only later by radiation from the hard sector), they can be expected to remain small in units of $Q_s$ throughout the entire evolution, due to dilution by the longitudinal expansion. Therefore the coupling to other gauge-invariant operators of the soft-sector of higher scaling dimensions will be suppressed by powers of $Q_s$. 


\subsection{The coupled equations of motion}

The minimalist approach to coupling the IR-CFT energy-momentum tensor $\mathcal{T}_{\mu\nu}$ to the weakly interacting glasma fields is a tensorial 
coupling\footnote{We will give a heuristic justification of this kind of coupling in the next section.} to $t_{\mu\nu}$, the energy-momentum tensor of the Yang--Mills fields. The latter has the standard expression 
\begin{equation}\label{tmn}
\st_{\mu\nu}(x) = \frac{1}{N_c} \text{Tr}\Big(F_{\mu\alpha}
F^{\phantom{\alpha}\alpha}_{\nu} - \frac{1}{4}\eta_{\mu\nu} F_{\alpha\beta}F^{\alpha\beta}\Big)\,,
\end{equation}
in matrix notation appropriate for the adjoint representation, where the factor $N_c$ has been introduced by
the trace normalization in the adjoint representation -- 
$\text{Tr}(T^aT^b)=N_c\delta^{ab}$, with $a,\,b=1\cdots N_c^2-1$ representing the color indices
for SU$(N_c)$ gauge group. We use the standard convention for the (gauge) covariant derivative: $D_\mu=\del_\mu-\rmi g A_\mu^aT^a$, where $g$ is the effective Yang-Mills coupling for the glasma fields as discussed above.

Furthermore, the scalar  operator $\mathcal{H}$  of the IR-CFT will naturally couple
to the `glueball' operator for the glasma fields, i.e. the density of the Yang--Mills action:
\beq
\sh(x)\,=\,\frac{1}{4N_c}
{\rm Tr}(F_{\alpha\beta}F^{\alpha\beta})\,.\eeq

Based on these considerations and dimensional arguments, we are led to the following minimalist
action describing the dynamics of the weakly--interacting glasma fields in the presence of the
strongly--interacting soft radiation:
\begin{eqnarray}\label{ac}
S_{\text{glasma}} =  -\int {\rm d}^4 x \, \frac{1}{4N_c}{\rm Tr}(F_{\alpha\beta}F^{\alpha\beta}) - \frac{\beta}{Q_s^4}\int {\rm d}^4 x\, h\,\mathcal{H} - \frac{\gamma}{Q_s^4}\int {\rm d}^4 x \,
\st_{\mu\nu}\mathcal{T}^{\mu\nu}.
\end{eqnarray}
Above, $\beta$ and $\gamma$ are two dimensionless parameters in this effective theory, which
at large $N_c$ must scale like $1/N_c^2$. In this case, all the terms in the above action
scale in the same way at large $N_c$, since the two IR--CFT operators $\mathcal{H}$ and
$\mathcal{T}_{\mu\nu}$ are expected to scale like $N_c^2$ in the large $N_c$ limit. On the other hand,
the expectation values of these IR-CFT operators are also expected to remain much smaller than $Q_s^4$
for reasons discussed above, hence the `hard--soft' couplings in \eqn{ac} should act
as a small perturbation on the dynamics of the hard modes. Also, with a slight abuse of notations, we have used the notations 
$\mathcal{H}$ and $\mathcal{T}_{\mu\nu}$ not only for the respective IR-CFT operators, but for the corresponding expectation values in the IR-CFT state to be determined self-consistently. We will continue to use these notations, it should be clear from the context whether we are referring to the IR-CFT operators or their expectation values.

The coupling to the IR-CFT modifies the equations for the glasma fields (inside the forward light--cone whose tip is at the event of collision; see also the discussion in
Sect.~\ref{sec:init} for more details) which can be readily obtained from  Eq. \eqref{ac} and read
\begin{eqnarray}\label{glback}
D_\mu F^{\mu\nu} =  -\frac{\beta}{Q_s^4} D_\mu (F^{\mu\nu}\mathcal{H})+\frac{\gamma}{Q_s^4} D_\mu (F^{\mu\nu}\mathcal{T}^{\alpha\beta}\eta_{\alpha\beta})
- \frac{2\gamma}{Q_s^4} D_\mu (\mathcal{T}^{\mu\alpha}
F_{\alpha}^{\phantom{\alpha}\nu} + F^{\mu}_{\phantom{\alpha}\alpha}
\mathcal{T}^{\alpha\nu}).
\end{eqnarray}

As announced, the dynamics of the strongly coupled IR-CFT will be studied using holographic methods.
Via the `hard--soft' couplings in \eqn{ac}, the glasma degrees of freedom act as sources for 
the IR-CFT, which is thus  \textit{marginally} deformed. In the dual, holographic, description, this implies that the 
dual fields in the bulk acquire non--trivial boundary values, so that the non-normalizable modes of these dual fields 
are non-vanishing. These boundary values  must equal to $(\beta/Q_s^4)\sh$ for the bulk dilaton
field (dual to the IR-CFT operator $\mathcal{H}$) and respectively  $(\gamma/Q_s^4)\st_{\mu\nu}$ for the bulk graviton  (dual to the IR-CFT operator $\mathcal{T}_{\mu\nu}$).

Let us denote the massless dilaton field in the bulk as $\Phi$ and the five-dimensional metric as 
$G_{MN}$. We use the Fefferman-Graham coordinates for the bulk spacetime,
where $x^M=(z,x^\mu)$, with $z$ denoting the holographic radial coordinate and $x^\mu$
the four field--theory coordinates. In these coordinates, the boundary of $AdS_5$ is
located at $z= 0$.
Then the near-boundary behavior of the bulk fields $\Phi$ and $G_{MN}$ reads
\begin{eqnarray}\label{bcs}
\Phi(z,x) &=&  \frac{\beta}{Q_s^4}\sh(x) + \dots + z^4 \frac{4\pi G_5}{l^3} \mathcal{H}(x) + \mathcal{O}\big(z^6\big), \nonumber\\
G_{rr}(z,x) &=& \frac{l^2}{z^2},\nonumber\\
G_{r\mu}(z,x) &=& 0,\nonumber\\
G_{\mu\nu}(z,x) &=& \frac{l^2}{z^2}\Big(\eta_{\mu\nu}+\frac{\gamma}{Q_s^4}\st_{\mu\nu}(x) + \dots + z^4 \left(\frac{4\pi G_5}{l^3}\mathcal{T}_{\mu\nu}(x) + P_{\mu\nu}(x)\right) \nonumber\\ &&\quad+ \mathcal{O}\big(z^4 \ln\, z\big)\Big),
\end{eqnarray}
where $\mathcal{H}$ and $\mathcal{T}_{\mu\nu}$ denote the respective expectation values in the dual IR-CFT state. 
The additional tensor $P_{\mu\nu}$ occurring in the coefficient of the $z^4$ term in the above expansion of $G_{\mu\nu}$ 
is a well-defined functional of the leading term $g_{(0)\mu\nu}\equiv \eta_{\mu\nu}+ (\gamma/Q_s^4) t_{\mu\nu}$,
which appears because this leading term itself (which, we recall, represents the metric in the dual IR-CFT theory) 
is different from the Minkowski metric $\eta_{\mu\nu}$ and hence it describes a curved space-time.
(We shall shortly return to a discussion of the physical meaning of this non-trivial boundary metric $g_{(0)\mu\nu}$.)
This tensor $P_{\mu\nu}$ can be constructed following the method described in Ref.~\cite{deHaro:2000xn}, which
yields
\begin{equation}
P_{\mu\nu} = \frac{1}{8}g_{(0)\mu\nu}\left(\left({\rm Tr}\, g_{(2)}\right)^2 - {\rm Tr}\, g_{(2)}^2\right) + \frac{1}{2}(g_{(2)}^2)_{\mu\nu}-\frac{1}{4}g_{(2)\mu\nu}{\rm Tr}\, g_{(2)}.
\end{equation}
where $g_{(2)\mu\nu}$ is the coefficient of $z^2$ in the expansion of $G_{\mu\nu}$, namely
\begin{equation}
g_{(2)\mu\nu} = \frac{1}{2}\left( R_{\mu\nu}[g_{(0)}] -\frac{1}{6} R[g_{(0)}]g_{(0)\mu\nu}\right).
\end{equation}
In the above expressions, all raising and lowering of indices have been defined with the metric $g_{(0)\mu\nu}$ or its inverse. Since, however, $g_{(0)\mu\nu}= \eta_{\mu\nu}+(\gamma/Q_s^4)t_{\mu\nu}$, one sees that $P_{\mu\nu}$ can be also viewed as a functional of $t_{\mu\nu}$ in the flat, Minkowski space-time.

Thus by expanding the self-consistently determined solution of the five-dimensional classical gravity equations, to be specified soon, in a power series of the radial coordinate near the boundary, one can obtain both  $\mathcal{H}$ and $\mathcal{T}_{\mu\nu}$. In \eqn{bcs}, $l$ is the `AdS radius' and determines the bulk five-dimensional cosmological constant via $\Lambda_5=-6/l^2$, and $G_5$ denotes the five-dimensional 
Newton's constant. Note that $l$ and $G_5$ do not appear separately in any physical 
quantity\footnote{This is more precisely true
in the present, classical gravity, approximation. The parameters $l$ and $G_5$ will appear separately 
if we consider corrections away from the strong coupling and large $N$ regime (giving higher derivative and quantum corrections to Einstein's theory in the bulk respectively).}, but only in the dimensionless combination $4\pi G_5/l^3$,
which according to the AdS/CFT dictionary should be identified with $2\pi^2/N_c^2$.
Recalling that both $\beta$ and $\gamma$ scale like $1/N_c^2$, we conclude that all the terms
shown in the expansion in \eqn{bcs} have the same scaling in the large $N_c$ limit: they are all of
$\order{1}$.

The five dimensional classical gravity theory giving equations for the dynamics of the massless bulk fields $\Phi$ and $G_{MN}$ can be taken to be those of Einstein's gravity with a minimally coupled massless dilaton and a negative cosmological constant, because a two-derivative action in the bulk is expected in the strong coupling limit. By also including a properly chosen dilaton potential $V(\Phi)$,  as e.g. in \cite{Gursoy:2007cb}, one could 
mimic the running of the QCD coupling in the `soft' sector. For simplicity, here we shall limit ourselves
to the case of a fixed coupling, so there is no potential for the dilaton. Then the bulk equations of motion 
take the standard form
\begin{eqnarray}\label{gravity}
R_{MN} - \frac{1}{2} R G_{MN} - \frac{6}{l^2} G_{MN} &=&  \nabla_M \Phi \nabla_N \Phi - \frac{1}{2}G_{MN}\nabla_P\Phi \nabla^P \Phi, \nonumber\\
G^{MN}\nabla_M\nabla_N\Phi &=& 0.
\end{eqnarray}
These equations must be solved with the boundary conditions as $z\to 0$ that can be read from 
Eq.~\eqref{bcs} together with initial conditions appropriate for the problem of heavy-ion collisions, to be
described in the next section. These constraints together with the requirement that there must
be a regular horizon in the bulk ensure the unicity of the solutions.

{In the standard holographic dictionary, the boundary metric, which is the leading term in the
near-boundary expansion of $G_{\mu\nu}$ in \eqn{bcs}, that is,
\begin{equation}\label{bmetric}
g_{(0)\mu\nu} (x)\,=\, \eta_{\mu\nu}+\frac{\gamma}{Q_s^4}\st_{\mu\nu}(x),
\end{equation}
is the metric in which the degrees of freedom of the boundary gauge theory `live'. In our case, 
this `boundary gauge theory' is of course the IR-CFT, which is not a fundamental theory by itself, 
but an effective theory for the strongly--coupled, soft, modes of the glasma. The fact that this
effective theory `lives' in a non-trivial metric should not be meant to imply that these 
non-perturbative degrees of freedom truly live in some curved space-time. Rather, it simply expresses
the fact that the IR-CFT is `deformed' by its coupling to the hard sector, meaning that 
hard and soft d.o.f. exchange energy with each other. As we now explain, this interpretation
naturally emerges from the Einstein equations in the bulk. Specifically, in the limit $z\to 0$, 
one of the constraints in the Einstein equations imply
the following conservation law for energy and momentum:
\begin{equation}\label{consv}
\nabla_\mu\mathcal{T}^{\mu\nu}(x)\,=\, \frac{\beta}{Q_s^4} \,\mathcal{H}(x)\nabla^\nu \sh(x),
\end{equation}
where the space-time indices are raised/lowered with the help of the boundary metric  \eqref{bmetric}
or its inverse, e.g.,
\begin{equation}\label{Tup}
\mathcal{T}^{\mu\nu} = g^{\mu\rho}_{(0)}\,\, \mathcal{T}_{\rho\sigma}\,\, g_{(0)}^{\sigma\nu}\,,
\end{equation}
and the covariant derivatives are written with respect to this metric as well.
Recalling that the Christoffel symbols for the Levi-Civita connection $\Gamma$ associated to this
metric are built with the energy--momentum tensor of the hard modes,
\begin{equation}
\Gamma^{\mu}_{\nu\rho}[\st] = \frac{\gamma}{2Q_s^4}\Big(\partial_\nu \st^\mu_{\phantom{\mu}\rho} + \partial_\rho \st^\mu_{\phantom{\mu}\nu} 
- \partial^\mu \st_{\nu\rho}\Big) + \mathcal{O}\big( \st^2\big),
\end{equation}
(in the r.h.s. of this equation, all lowering and raising of indices are done with respect to the flat metric 
$\eta_{\mu\nu}$ and its inverse), one sees that the conservation law in \eqn{consv}
can be rewritten in terms of 
ordinary derivatives in flat, Minkowski, space-time and can be properly interpreted as the standard 
conservation of energy and momentum, but in the presence of an external `driving force'
representing the hard degrees of freedom; that is,
 \begin{eqnarray}\label{consv2}
\partial_\mu\mathcal{T}^{\mu\nu} \, &=&\,  \frac{\beta}{Q_s^4}\, \mathcal{H} \,g_{(0)}^{\mu\nu}[t]\,
\partial_\mu \sh \,
 - 
\Gamma^\gamma_{\alpha\gamma}[\st]
\mathcal{T}^{\alpha\nu} -
\Gamma^\nu_{\alpha\beta}[\st]
\mathcal{T}^{\alpha\beta}.
\end{eqnarray}
This is precisely the expected conservation law for a generic field theory (here, our IR-CFT) which
is `deformed' by the irrelevant operators shown in \eqn{ac}. 
Independently of this holographic context,
this law can also be obtained as a Ward identity via standard manipulations on the path integral for the IR-CFT.
Notice that, although exact (in an operator sense), this equation \eqref{consv2} is not sufficient
to determine the dynamics of the energy--momentum tensor in the IR-CFT (for given hard `sources'
$\sh$ and $\st_{\mu\nu}$), 
because there are only 4 such equations constraining 10 independent degrees of freedom.
As a matter of fact,  there is still another constraint --- an equation for the trace of $\mathcal{T}^{\mu\nu}$ 
as computed in the boundary metric, which should be reinterpreted too as an equation in flat space 
involving hard `sources' ---, but the number of such constraints is still too low to fully fix the dynamics,
as expected. One needs to solve the complete bulk dynamics, meaning \eqref{gravity}, to obtain the evolution of $\mathcal{T}^{\mu\nu}$ and $\mathcal{H}$.

The previous discussion also points towards a rather subtle point, that we shall now try to elucidate:
since, as we have just seen, the IR-CFT sector formally lives in a curved space-time with metric $g_{(0)\mu\nu} (x)$,
whereas the Yang-Mills sector lives in the physical space-time with Minkowski metric 
$\eta_{\mu\nu}$, there is {\em a priori} an ambiguity in the construction of the tensorial coupling between the 
energy-momentum tensors of these two sectors. In writing the action \eqref{ac}, we have chosen this coupling 
in the form $t_{\mu\nu}\mathcal{T}^{\mu\nu}$, which is of course\footnote{Needless to say, all the equations
referring to the hard glasma fields, like Eqs.~\eqref{tmn}--\eqref{glback}, are written in Minkowski metric. For these equations to be manifestly Lorentz-covariant, one should use the
Minkowski tensor  $\tilde{\mathcal{T}}_{\mu\nu}$ which is defined by  $\tilde{\mathcal{T}}^{\mu\nu}
\equiv \mathcal{T}^{\mu\nu}$. In practice, and in order to avoid possible confusion,
we payed attention to explicitly use the contravariant components in writing 
Eqs.~\eqref{ac} and \eqref{glback}.}
the same as $t^{\mu\nu}\tilde{\mathcal{T}}_{\mu\nu}$, with $\tilde{\mathcal{T}}_{\mu\nu}
\equiv \eta_{\mu\alpha}\eta_{\nu\beta}\mathcal{T}^{\alpha\beta}$, but is {\em not} the same
as $t^{\mu\nu}\mathcal{T}_{\mu\nu}$, with $\mathcal{T}_{\mu\nu}$ the covariant tensor
which enters the near-boundary expansion of $G_{\mu\nu}$ shown in \eqref{bcs}. (The relation between
the covariant
and contravariant components of this tensor is rather given by \eqn{Tup}.) Our above choice to minimally
couple to contravariant tensor $\mathcal{T}^{\mu\nu}$ to the corresponding tensor 
of the hard modes --- that is, to define the hard-soft coupling as $t_{\mu\nu}\mathcal{T}^{\mu\nu}$,
rather than $t^{\mu\nu}\mathcal{T}_{\mu\nu}$ --- is truly a prescription, which should be viewed as a part 
of our model and is ultimately guided by physical considerations: from the viewpoint of the IR-CFT theory,
the `source' $t_{\mu\nu}$ plays the same role as a deformation $\delta g_{\mu\nu}$ in the metric, and hence
it is most naturally considered with covariant indices.

To conclude, the glasma fields $F_{\mu\nu}^a$ and the
expectation values $\mathcal{H}$ and $\mathcal{T}^{\mu\nu}$ of the IR-CFT operators must be simultaneously  determined by self--consistently solving the 
generalized Yang--Mills equations \eqref{glback} and the classical gravity equations \eqref{gravity},
with the boundary conditions \eqref{bcs} for the latter. 
For the corresponding solutions to be uniquely determined, we still need to specify the
initial conditions at $\tau=0$ and the regularity condition in the bulk. These additional ingredients will be discussed
in the next two subsections, together with more details on the numerical implementation of the
self--consistency procedure.

It is natural to associate the hydrodynamic variables at late time with the energy-momentum tensor of the soft modes, namely $\mathcal{T}^{\mu\nu}$. Indeed, one expects these soft modes to be the first ones to
thermalize. The hydrodynamics of these soft modes however should be formulated in the deformed metric 
\eqref{bmetric}, which `knows' about the hard fields via the respective energy-momentum tensor $t_{\mu\nu}$.
This leads to conservation laws like \eqn{consv2}, which truly describe {\em forced}
hydrodynamics in flat, Minkowski, space-time, with the forcing terms generated by the hard modes. At late time, the dynamics of the hard modes can be given by a suitable Boltzmann equation with a non-standard collisional kernel. (We will return to this point in the concluding section.) Here we simply stress that even the late time dynamics in our scenario is expected to be richer than in more conventional approaches.

Notice that the time scale for thermalization and the onset of a hydrodynamic description 
will be affected by both the glasma and the holographic degrees of freedom. So, our 
scenario cannot be expected to predict universal values for the thermalization time 
and the transport coefficients, as is generally the case for the purely holographic models at strong coupling. (In particular, the non--linear dynamics of the glasma fields can by itself lead to fast isotropization \cite{Gelis:2013rba}.)
All observables will be determined in terms of the two free parameters of the model, 
namely the couplings $\beta$ and $\gamma$ in the action \eqref{ac}, together with the dimensionless quantity
$4\pi G_5/l^3=2\pi^2/N_c^2$ of the classical gravity theory in the bulk. 



\subsection{Initial conditions and the iterative algorithm}
\label{sec:init}

A main virtue of our present semi--holographic model is that it provides an unambiguous  
initial--condition problem (including for the holographic sector), which is entirely determined by 
the pQCD picture for the colliding nuclei --- more precisely, by the associated CGC--description
in terms of color charge distributions \cite{Iancu:2002xk,Gelis:2010nm} 
(see also below). This should be contrasted to purely holographic
approaches, which requires an ad-hoc initial condition for the gravitational problem, that
cannot be directly related to properties of the colliding nuclei. 

It is first of all clear that the initial conditions in the hard sector, i.e. for the classical glasma fields,
will not be affected by the holographic degrees of freedom. Using causality and the uncertainty principle, 
one can argue that the expectation values of
$\mathcal{T}_{\mu\nu}$ and $\mathcal{H}$ (the soft sector operators) must vanish at proper time $\tau = 0$
(the time where the collision is initiated) : indeed, it takes some time $\tau\sim 1/p_\perp \gtrsim 1/Q_s$ to emit soft gluons with transverse momenta $p_\perp\lesssim Q_s$. Hence, when $\tau\to 0$, 
the modified glasma equations \eqref{glback} reduce to the standard Yang--Mills equations and  
the initial conditions for our glasma fields are the same as
in the standard CGC approach at weak coupling. These initial conditions have been worked out in
\cite{Kovner:1995ja, Kovner:1995ts} and will be briefly discussed below.

Before revising the standard CGC initial conditions for the glasma fields, let us also consider the corresponding
problem on the gravitational side. As is well known, the gravitational problem in an asymptotically $AdS_5$ spacetime cannot be reduced to an initial--value problem alone. Rather, one also need to specify the boundary ($z\to 0$) 
metric and the boundary value of the dilaton field at 
\emph{all} times --- to ensure a well-defined evolution that leads to a space--time with a smooth future horizon. 
In our case, these boundary data are specified by \eqref{bcs} in terms of the glasma fields. However, in order
to know the latter at all times, we need to solve the modified glasma equations \eqref{glback}, which in turn
involve the `soft' operators $\mathcal{T}_{\mu\nu}$ and $\mathcal{H}$, as determined by the solutions to the
gravitational problem. Hence, we are facing a self--consistent problem, whose solutions 
calls for an iterative procedure. 
In what follows, we will lay out a well-defined procedure in that sense.


Let us first review the standard CGC initial conditions for the glasma fields \cite{Kovner:1995ja, Kovner:1995ts}. 
To that aim, we recall the standard glasma equations, which do not include the soft degrees of freedom,
but where the color sources representing the colliding nuclei are manifest\footnote{These
color sources are not  apparent in \eqn{glback} since their supports are restricted to the two
light--cones $x^3=\pm t$, whereas  
\eqn{glback} is written for $t>|x^3| >0$, i.e. inside
the forward light--cone.}:
\begin{eqnarray}\label{YMrho}
D_\mu F^{\mu\nu}(x)\, =\,\delta^{\nu+}\rho_{(1)}(x^-,\bm{x}_\perp)\,+\,
\delta^{\nu-}\rho_{(2)}(x^+,\bm{x}_\perp)\,.
\end{eqnarray}
We are using light--cone coordinates: $x^\mu=(x^+, x^-, \bm{x}_\perp)$,
with $x^\pm=(t\pm x^3)/\sqrt{2}$ and $\bm{x}_\perp=(x^1,x^2)$.
Here, $J^{\nu\,a}_{(1)}(x)=\delta^{\nu+}\rho_{(1)}^a(x^-,\bm{x}_\perp)$ is the color current of the
first nucleus, which is a `right mover' (it propagates towards increasing $x^3$ at nearly the speed of light). 
The corresponding color charge density  $\rho_{(1)}^a(x^-,\bm{x}_\perp)$ is independent of $x^+$
by Lorentz time dilation and is localized near $x^-=0$ by Lorentz contraction; roughly,
$\rho_{(1)}^a(x^-,\bm{x}_\perp)\approx \delta(x^-)\rho_{(1)}^a(\bm{x}_\perp)$. A similar
discussion, with $x^+\leftrightarrow x^-$, applies to the second nucleus, which is a left mover
and has a color current
$J^{\nu\,a}_{(2)}(x)=\delta^{\nu-}\rho_{(2)}^a(x^+,\bm{x}_\perp)$. 
The color charge densities are randomly distributed, independently for each nucleus, 
according to a gauge--invariant probability distribution (the `CGC weight function'),
which in the simplest approximation (known as the `McLerran--Venugopalan model') is
taken to be a Gaussian:
\begin{equation}\label{gauss}
\langle \rho^{a}_{(m)} (\bm{x}_\perp)\rho^{b}_{(n)} (\bm{y}_\perp)\rangle = g^2\mu^2
\,\delta_{mn} \,\delta^{ab}\, \delta^{(2)}(\bm{x}_\perp-\bm{y}_\perp).
\end{equation}
In this equation, the indices $m,\,n=1,\,2$ refer to the two nuclei 
and the 2--dimensional density of color charge squared $g^2\mu^2$ is 
proportional to the saturation scale $Q_s^2$ \cite{McLerran:1993ni, McLerran:1993ka, McLerran:1994vd}.

\eqn{YMrho} is written in the Schwinger gauge $x^+A^-+x^-A^+=0$, in which the current
$J^\nu=J^{\nu}_{(1)}+J^{\nu}_{(2)}$ is not rotated by the classical field\footnote{In a generic gauge,
the current should be dressed with Wilson lines to maintain its covariant conservation.}.
The solution to this equation outside the forward light--cone 
--- meaning prior to the collision ($t<0$) and in the region which is causally disconnected
from the collision at $t>0$ --- is the superposition
of the individual fields of the two nuclei, which outside 
the light cones are 2--dimensional pure gauges :
 \begin{align}
 A^+\,&=\,A^-\,=\,0\,,\nn
 A^i(x)&=\theta(-x^+)\theta(x^-)A^i_{(1)}(\bm{x}_\perp) +\theta(-x^-)\theta(x^+) A^i_{(2)}(\bm{x}_\perp),
 \end{align}
 where the upper index $i=1,2$ denotes the transverse component and
\begin{eqnarray}\label{glin2}
A_{(1,2)}^i(\bm{x}_\perp) &=& \frac{\rmi}{g}\, U_{(1,2)}(\bm{x}_\perp)\partial_i U^\dagger_{(1,2)}(\bm{x}_\perp), \nonumber\\
U_{(1,2)}(\bm{x}_\perp) &=& {\rm P \, \, exp}\Bigg(- \rmi g
 \int {\rm d}x^{\mp}\frac{1}{\nabla_\perp^2}\rho_{(1,2)}(x^\mp, \bm{x}_\perp) \Bigg)\,,
\end{eqnarray}   
with the $P$ symbol indicating the path
ordering of the color matrices $\rho_{(1,2)}=\rho_{(1,2)}^aT^a$ in the exponent w.r.t. the
relevant longitudinal coordinate, $x^-$ or $x^+$.

After the collision, 
the solution to \eqn{YMrho} in the future light cone ($t>|x^3|$)
is {\em boost--invariant}, meaning that it depends upon the proper--time $\tau$,
but not on the rapidity $\eta$. These variables $\tau$ and $\eta$ are defined
as (for $t > |x^3|$)
\beq\label{taueta}
\tau\,=\,\sqrt{t^2-(x^3)^2}\,=\,\sqrt{2x^+x^-}\,,\qquad \eta\,=\,\frac{1}{2}\ln\frac{x^+}{x^-}\,
.\eeq
More precisely, the solution (in the gauge $A^{\tau}=(x^+A^-+x^-A^+)/\tau = 0$) 
takes the form $A^\pm(x)=\pm x^\pm A^\eta(\tau,\bm{x}_\perp)$
and $A^i(x)=A^i(\tau,\bm{x}_\perp)$, with the functions $A^\eta$ and $A^i$ 
obeying the following initial
conditions at $\tau=0$ :
\begin{align}\label{glin1}
A^i = A^i_{(1)} +A^1_{(2)}, \qquad A^\eta =  \frac{\rmi g}{2}\big[A^i_{(1)}, A^i_{(2)}\big], \qquad
\partial_\tau A^i = 
\partial_\tau A^\eta = 0,
\end{align}
where the square brackets refer to the commutator of color matrices.

As previously mentioned, in our semi--holographic approach, the initial conditions \eqref{glin1} also
applies to the complete equation of motion for the glasma fields, i.e. \eqn{glback} which includes the
back reaction of the soft modes. For this equation to be fully specified though, we also need the 
expectation values $\mathcal{H}$ and $\mathcal{T}_{\mu\nu}$ of the IR-CFT operators, which in turn
are determined by solving the gravitational equations with boundary conditions specified by
the glasma fields, cf. \eqn{bcs}. As discussed earlier, this gravitational problem is not just an initial--value problem:
it requires specifying the boundary data at {\em all} times. To cope with this, we propose a
self--consistent procedure in which the back reaction of the soft sector on the glasma fields is
formally treated as a `small perturbation', which is taken into account via successive iterations
--- to be repeated until one achieves convergence. 

Specifically, in the first iteration, one solves the glasma equation without the holographic d.o.f. in the forward light--cone, that is the homogeneous version of 
\eqn{YMrho} with the initial conditions \eqref{glin1}. This numerical solution, which is well known in the literature
\cite{Krasnitz:2001qu,Lappi:2003bi}, 
is then used to compute the boundary conditions for the classical gravity fields, at all times.
This requires performing the average over the distribution \eqref{gauss} of the color charge
densities in the colliding nuclei, which in practice is done by randomly selecting the initial color charge
densities according to the Gaussian distribution  \eqref{gauss}.
Strictly speaking, the quantities $\st_{\mu\nu}$ and $\sh$ thus obtained exhibit logarithmic divergences
as $\tau\to 0$ and require an ultraviolet cutoff. However, as argued in Ref.~\cite{Lappi:2006hq}, this divergence
has no influence on the soft sector because $\st_{\mu\nu}$ and $\sh$ become independent of the UV cutoff
at later times $\tau\gtrsim 1/Q_s$, which is when we actually need them (since, as previously explained,
it takes some finite time $\tau\gtrsim 1/Q_s$ to emit the soft fields).
To avoid such complications with the limit $\tau\to 0$ in the 
numerical simulations, we propose to smoothly switch on the interactions between
the hard and soft sectors, by replacing the previous couplings $\beta$ and $\gamma$
in \eqn{ac} with their time--dependent versions
\begin{eqnarray}\label{td}
\beta= \beta_0 \tanh \big(\alpha_0\tau Q_s\big), \qquad
\gamma= \gamma_0 \tanh\big(\alpha_0\tau Q_s\big),
\end{eqnarray}
where $\alpha_0$, $\beta_0$ and $\gamma_0$ are new, dimensionless parameters. This 
in particular implies that the boundary sources  for the gravitational problem will acquire a similar time-dependence:
\begin{eqnarray}\label{sc}
\lim_{z\rightarrow 0}\frac{z^2}{l^2}G_{\mu\nu} = g_{\mu\nu}^{({\rm b})} &=& \eta_{\mu\nu} + \frac{\gamma_0}{Q_s^4} \tanh
 \big(\alpha_0\tau Q_s\big)\st_{\mu\nu},\nonumber\\
\lim_{z\rightarrow 0}\Phi  &= & \frac{\beta_0}{Q_s^4}\tanh \big(\alpha_0\tau Q_s\big)\sh,
\end{eqnarray}
at the boundary $r = 0$. Similarly the modified glasma equations \eqref{glback} should now read
\begin{eqnarray}\label{glback2}
D_\mu F^{\mu\nu} &=&  -\frac{\beta_0}{Q_s^4} D_\mu \left(\tanh \big(\alpha_0\tau Q_s\big)F^{\mu\nu}\mathcal{H}\right)+\frac{\gamma_0}{Q_s^4} D_\mu \left(\tanh \big(\alpha_0\tau Q_s\big)F^{\mu\nu}\mathcal{T}^{\alpha\beta}\eta_{\alpha\beta}\right)
\nonumber\\ &&- \frac{2\gamma_0}{Q_s^4} D_\mu \left(\tanh \big(\alpha_0\tau Q_s\big)\left(\mathcal{T}^{\mu\alpha}
F_\alpha^{\phantom{\alpha}\nu} + F^{\mu}_{\phantom{\mu}\alpha}
\mathcal{T}^{\alpha\nu}\right)\right).
\end{eqnarray}
As clear from \eqref{sc}, on the gravitational side, due to vanishing of the hard--soft couplings at the initial time $\tau=0^+$, the boundary metric is just the flat metric $\eta_{\mu\nu}$ and the boundary value of the dilaton field is vanishing at the initial time. Therefore, we can consistently put the initial conditions on the gravity side to be the vacuum pure $AdS_5$ solution with a vanishing dilaton. 

Following \cite{Chesler:2013lia}, we shall employ the ingoing Eddington-Finkelstein coordinates, in which the bulk space--time metric can be written as
\begin{eqnarray}\label{bmetricEF}
{\rm d}s^2 = G_{mn}(X) {\rm d}x^m {\rm d}x^n + 2{\rm d}\tau \left({\rm d}r - A(X) {\rm d}\tau - F_m(X){\rm d}x^m\right),
\end{eqnarray}
where $X$ denotes collectively all bulk coordinates, $r$ is the new radial coordinate\footnote{The relation between the radial Fefferman--Graham coordinate $z$ and the radial Eddington--Finkelstein coordinate $r$ is $r= l^2/z + \mathcal{O}(z^0)$ near the boundary. The relation between the two time coordinates $\tau_{{\rm FG}}$ and $\tau_{{\rm EF}}$ is $\tau_{{\rm EF}} = \tau_{{\rm FG}} - z +O(z^2)$. All the field-theory coordinates $x^\mu$ in both these coordinate systems agree at the boundary but they differ in the interior (see \cite{Gupta:2008th} for instance).  For simplicity, we do not use different notations for the field-theory coordinates in these two coordinate systems.}, $\tau$ is the proper time defined in \eqref{taueta}, and $x^m$ denotes the three spatial coordinates, namely, the `rapidity coordinate' $\eta$ defined in \eqref{taueta} and the transverse coordinates $x^i$ with $i=1,\,2$. In these coordinates, the boundary is at $r= \infty$. The components of the boundary metric $g_{\mu\nu}^{(\rm b)}$ as written in
\eqref{sc} are related to the boundary limit of the components of $G_{\mu\nu}$ of the new metric via
\begin{equation}
g_{mn}^{({\rm b})} = \lim_{r\to \infty}\frac{l^2}{r^2}G_{mn}, \quad g_{\tau\tau}^{({\rm b})} = -2\lim_{r\to \infty}\frac{l^2}{r^2}A, \quad g_{\tau m}^{({\rm b})} = -\lim_{r\to \infty}\frac{l^2}{r^2}F_m.
\end{equation}
Similarly in the second equation of \eqref{sc}, the right hand side refers now to the limit $r\to\infty$ of the dilaton field $\Phi$ in the new coordinates.
  
In order to ensure the ``vacuum initial conditions'' for the soft sector, we will require the following
initial conditions at $\tau = 0^+$ \footnote{We recall that the components $\mathcal{T}_{\mu\nu}$ 
specify the subleading term, of order $z^4$, in the near-boundary
 expansion of the metric $G_{\mu\nu}$ in Fefferman-Graham  coordinates, cf. \eqn{bcs}. The components of $\mathcal{T}_{\mu\nu}$ can also be read from a similar asymptotic expansion of the Eddington-Finkelstein coordinates (see \cite{Chesler:2013lia} for instance).}
\begin{eqnarray}\label{gravin}
G_{\eta\eta}(r, \eta, x^i) = \frac{r^2}{l^2}\tau^2,\,\, G_{ij}(r, \eta, x^i) =\frac{r^2}{l^2} \delta_{ij}, \,\, G_{\eta i}(r, \eta, x^i) = 0, \nonumber\\
\mathcal{T}_{\tau\tau}(\eta, x^i) = \mathcal{T}_{\tau\eta}(\eta, x^i) = \mathcal{T}_{\tau i}(\eta, x^i) = 0,\nonumber\\
\Phi(r, \eta, x^i) = \partial_\tau \Phi (r, \eta, x^i)\,\, = \, \, 0.
\end{eqnarray}
Notice that we need to specify the spatial components $G_{mn}$ {\em entirely} (meaning for all values of $r$), while for the functions $A$ and $F_m$ we need only the `subleading normalizable modes', which are $\mathcal{T}_{\tau\tau}$ and $\mathcal{T}_{\tau m}$ respectively\footnote{The fact that we can formulate the initial data directly in Eddington-Finkelstein coordinates allows us to remove any residual gauge degree of freedom and the obtain the rather simple form of the metric shown in \eqn{gravin}. In other approaches, where gravitational shock waves are collided, the initial data 
are naturally formulated (in analytic form) only in Fefferman-Graham coordinates. In such a case, the rewriting of
the initial data  in Eddington-Finkelstein coordinates requires to numerically perform a diffeomorphism, which can
be the source of residual gauge freedom already at the initial time \cite{Kinoshita:2008dq,Chesler:2013lia}. Fortunately,  this additional complication is not relevant to us.}.  This is because $A$ is an auxiliary variable whose
time--derivatives do not appear in the classical gravity equations.  Furthermore, after separating the classical gravity equations into equations for the dynamical evolution and constraints, it follows that time derivatives of $F_m$ also do not appear in the actual time evolution equations (see \cite{Chesler:2013lia} for more details). The radial equations for $A$ and $F_m$ can be integrated with the boundary conditions shown in \eqref{sc} and 
\eqref{gravin}\footnote{To be more precise, in this formulation ${\rm det}\, G_{mn}$ is also an auxiliary variable -- therefore we need to know $\hat{G}_{mn} = G_{mn}/({\rm det}\, G)$ only at the initial time for all values of $r$.}. The constraints of classical gravity equations, namely \eqref{consv}, give the time evolution of $T_{\tau\tau}$ and $T_{\tau m}$. 

    The Chesler-Yaffe algorithm \cite{Chesler:2013lia} can then be employed to solve the gravitational equations \eqref{gravity} with the initial conditions and boundary conditions specified as above. This algorithm ensures that the solution is smooth at the future horizon thus protecting causality in the dual field theory. The key to obtain the regularity at the future event horizon is to enforce an apparent horizon in the bulk geometry at a fixed value of the radial coordinate, which in our case will be put at $r= l^2\Lambda_{{\rm QCD}}$ at all times. We will discuss more about this in the next subsection. This regularity condition fixes the values of the remaining integration constants, thus fully specifying the solutions to the classical gravity equations.

From the above regular solution of gravity, we can extract $\mathcal{T}_{\mu\nu}$ (thus $\mathcal{T}^{\mu\nu}$ via \eqref{Tup})  and $\mathcal{H}$ by doing the near-boundary expansion of $G_{MN}$ and $\Phi$ respectively (recall \eqref{bcs}). We can then use it in \eqref{glback2} to correct the solution for the glasma fields in the next step of the iteration. The corresponding glasma solution should be used to recalculate 
the sources at the boundary \eqref{sc} in the gravitational problem, leading to new expectation values for the IR-CFT operators. As announced, this iterative procedure should be repeated until we obtain convergence for all
quantities (hard and soft) and at all times.

The iterative procedure is summarized below.
\begin{itemize}
\item In the first step of the iteration, we solve the glasma equations  \eqref{glback2} on the forward light cone with the initial conditions \eqref{glin1} and the expectation values of the soft-sector operators $\mathcal{T}^{\mu\nu}$ and $\mathcal{H}$ set to zero. 

\item The numerical solution of the glasma equations is then used to evaluate the sources \eqref{sc} in the gravitational problem.
\item With the above boundary sources and the initial conditions \eqref{gravin}, the gravitational equations \eqref{gravity} are solved numerically using the Chesler-Yaffe procedure ensuring that the solution is regular at the future horizon.
\item The gravitational solution is then expanded near the boundary ($z=0$ in Fefferman--Graham coordinates or $r=\infty$ in Eddington--Finkelstein coordinates) in order to extract the time-dependent expectation values of $\mathcal{T}^{\mu\nu}$ and $\mathcal{H}$.
\item The above are then used in the glasma equations of motion \eqref{glback2} to correct for the solution of the glasma fields obtained in the first iteration. The same initial conditions \eqref{glin1} are used to obtain the new solution for the glasma fields.
\item The corrected glasma solution is then used to recalculate the boundary sources \eqref{sc} in the gravitational problem.
\item The gravitational equations for the bulk geometry and bulk dilaton are then solved again with the above boundary conditions and the same initial conditions \eqref{gravin}.
\item The corrected time-dependent expectation values of $\mathcal{T}^{\mu\nu}$ and $\mathcal{H}$ are then extracted from the gravitational solution and used as inputs to correct the glasma solution. 
\item The iteration between the solution for the glasma fields and the solution for the gravitational equations continues until we reach convergence for both. Thus we obtain the final solution for both hard and soft fields.
\end{itemize}

We emphasize that the initial conditions for the glasma fields (namely \eqref{glin1}) and for the bulk metric and the dilaton (namely \eqref{gravin}) remain the same at every step in the iteration. Accordingly, the only physical
ingredient in fixing the initial value problem is the distribution of the color sources in the colliding 
nuclei, as given by the MV model (or, more generally, by the CGC effective theory \cite{Iancu:2002xk,Gelis:2010nm}).
As previously noticed, both the glasma equations and gravitational equations with similar degree of complexity
have already been solved in the past, via numerical methods. Therefore, it seems that our algorithmic procedure can be readily implemented with currently available numerical codes. It remains to be seen if, in practice, the
convergence of our iterative procedure can be achieved fast enough to allow for explicit
studies of thermalization and for phenomenological predictions.

\subsection{The apparent horizon and hadronization}
\label{sec:reg}

The Chesler-Yaffe algorithm  \cite{Chesler:2013lia} requires cutting out the gravitational metric behind the apparent horizon, which by exploitation of a residual gauge symmetry is fixed to a constant value of the radial coordinate. This residual gauge symmetry is the freedom of changing the radial coordinate from $r$ to $r'$ via $r =r' + \lambda(\tau, \eta, x^i)$, which keeps the form of the metric \eqref{bmetricEF} invariant. As $\lambda$ does not depend on the radial coordinate, this freedom can be always utilized to keep the apparent horizon at a fixed value of the radial coordinate. In practice this makes the variable $\lambda$ dynamical and also requires us to provide its value at the initial time. The natural initial value for $\lambda$ in our case is $\lambda = 0$ as this makes the spacetime metric \eqref{bmetricEF} match to a pure AdS metric on a standard constant time slice, at the initial time.

As the actual event horizon is `above' the apparent horizon (meaning closer to the boundary), this is sufficient to ensure that the solutions for both the bulk metric and the dilaton are smooth at the event horizon, thus satisfying the required regularity condition. The formation of the event horizon in the bulk signifies entropy production and then thermalization to an expanding perfect-fluid like state dual to an accelerating black hole \cite{Janik:2005zt}.

In our case, the apparent horizon has a specific physical significance by itself. The radial coordinate acts as the scale of a renormalization group flow (see next Section for more details), thus we
can think of the apparent horizon as a (geometric) infrared regulator for the soft modes. Indeed, the QCD modes with momenta below $\Lambda_{{\rm QCD}}$ (the confinement scale)
are not correctly accounted for by our IR-CFT and its holographic dual. To that aim, we shall 
demand that in some precise set of coordinates we should get an apparent horizon in the bulk at $r = l^2 \Lambda_{{\rm QCD}}$.  
This will give a concrete realization of the expectation that degrees of freedom below the confinement scale
should play no role in the dynamics leading to thermalization.

The effective temperature at late time can be read off by matching the late time behavior of $\mathcal{T}_{\mu\nu}$ to a hydrodynamic asymptotic series expansion in the proper time $\tau$ \cite{Janik:2005zt}.
At later times, the fireball will cool down due to expansion, meaning that the effective temperature will drop below $\Lambda_{{\rm QCD}}$.
This time will define the time of hadronization. 
The latter can be modeled to in our approach by changing the apparent horizon $r = l^2\Lambda_{{\rm QCD}}$ into a hard wall cut-off immediately when the effective temperature equals $\Lambda_{{\rm QCD}}$. The hard wall cut-off denotes boundary conditions which render the bulk Hamiltonian Hermitean. This has previously been used as an $AdS$ model of confinement in QCD  \cite{Polchinski:2001tt,BoschiFilho:2002vd,Erlich:2005qh}. The bulk excitations of both the metric and the dilaton can be decomposed into eigenfunctions each of which represent specific hadronic excitations\footnote{To be more realistic we need to also add gauge fields in the bulk which will capture the hadronic excitations with flavour quantum numbers \cite{Erlich:2005qh}. These will be also needed to understand the role of chemical potentials in the thermalization process. We can ignore these complications because the chemical potentials are expected to be small.}. This will be legitimate because the glasma fields will have dissipated into soft excitations and will be negligibly small -- and thus will play no role in the hadronization process.

\section{A heuristic derivation using renormalization group flow}

It is generally admitted that the radial dimension in the gravity set-up somehow captures the scale of a special renormalization group (RG) flow in the dual field theory, however the way how this actually happens is one of the outstanding problems regarding a fundamental understanding of the AdS/CFT correspondence. 

One can think of an abstract definition of the strong interaction limit as opening of a large gap in the scaling dimensions of the operators\footnote{This has to be understood in the sense of the parametric dependence of the anomalous dimensions of the various operators upon a field-theory coupling, such as the `t Hooft coupling in Yang-Mills theory. For most operators in the theory, the anomalous dimensions will receive parametrically large quantum corrections, therefore at infinitely strong coupling the corresponding operators decouple. The operators which survive in the strong coupling limit are those which are protected by symmetries and for which the scaling dimensions remain bounded even when the coupling becomes arbitrarily large. }.
Furthermore, the operators which survive this limit form an algebra generated by a few,
\emph{single-trace}, operators, namely the energy-momentum tensor $\mathcal{T}^{\mu\nu}$, conserved currents $\mathcal{J}_\mu$, and relevant condensates $\mathcal{H}$. All the other operators in the strong interaction limit are 
multi-trace operators like $\mathcal{T}^{\mu\nu}\mathcal{T}^{\rho\sigma}$.

Another pre-requisite for the field theory to have a holographic classical gravity dual, is the presence of a parameter $N$, such that in the large $N$ limit, the expectation values of multi-trace operators factorize, e.g.
\begin{equation}
\langle\mathcal{T}^{\mu\nu}\mathcal{T}^{\rho\sigma}
\rangle = \langle\mathcal{T}^{\mu\nu}
\rangle\langle\mathcal{T}^{\rho\sigma}\rangle
+ \mathcal{O}\left(\frac{1}{N^2}\right),
\end{equation}
where the expectation value refers to any state of the theory. We note that, despite this factorization property, the single trace operators will mix with multi-trace operators under the RG flow, thus giving rise to non-linearities. This is the fundamental reason why the dual classical gravity theory is non-linear.

Let us give here a general argument, to be developed in detail in a further publication \cite{Behr:2015yna},
about how the classical gravity theory can emerge from the dual field theory via a special renormalization group
(RG) flow.
We shall first discuss the pure holographic set-up and then generalize to the case of semi-holography.
Our argument will be restricted to the hydrodynamic sector describing the dynamics on large space-time scales,
where the evolution is governed by local conservation laws.
Under this assumption, we would like to argue that the gauge/gravity correspondence can be understood as a reconstruction of a special, {\em highly efficient}, RG flow, which can only exist in the large $N$ limit
(a condition which is necessary but in general not sufficient).


Consider projecting the degrees of freedom in the field theory on a certain slowly varying subsector via a projection operator $\mathcal{P}(\Lambda)$, where $\Lambda$ labels the typical scale for space-time variation for fields within this subsector. We also consider the following projection of an observable like the energy-momentum tensor into that subsector:
\begin{equation}
\mathcal{T}^{\mu\nu}(\Lambda) = \mathcal{P}(\Lambda)\mathcal{T}^{\mu\nu}\mathcal{P}^{\dagger}(\Lambda).
\end{equation} 
The result of such a projection is a scale-dependent, coarse-grained, observable. This observable will not satisfy the usual microscopic Heisenberg equation of motion.  For instance, unlike the microscopic energy-momentum tensor which satisfies the standard conservation equation $\partial_\mu \mathcal{T}^{\mu\nu} = 0$, its coarse-grained version $\mathcal{T}^{\mu\nu}(\Lambda)$ will satisfy a more complicated equation
\begin{equation}\label{comp}
\partial_\mu \mathcal{T}^{\mu\nu}(\Lambda) = \text{a non-linear functional of $\mathcal{T}^{\mu\nu}(\Lambda)$,}
\end{equation}
reflecting the fact that the hard degrees of freedom that have been projected out will generate driving forces for the slow moving sector. For simplicity, we have assumed that $\mathcal{T}^{\mu\nu}$ is the only single-trace operator. The non-linear contributions on the right hand side are due to the mixing of $\mathcal{T}^{\mu\nu}(\Lambda)$ with the multi-trace operators $(\mathcal{T}^{\mu\nu}
\mathcal{T}^{\rho\sigma})(\Lambda)$, etc. under the RG flow.

The main point for constructing a ``highly efficient RG flow'' is to construct a sequence of special projection operators $\mathcal{P}(\Lambda)$ such that the above equation takes a very special form, namely}
\begin{eqnarray}\label{compspecial}
\partial_\mu \mathcal{T}^{\mu\nu}(\Lambda) &=&\frac{1}{4\Lambda^4}\partial^\nu\left(
\mathcal{T}^{\alpha\beta}(\Lambda)\,\,\mathcal{T}^{\gamma\delta}(\Lambda)\,\,\eta_{\alpha\gamma}\,\,\eta_{\beta\delta}\right) 
-\frac{1}{2\Lambda^4}\left(\partial_\alpha {\rm Tr}\, \mathcal{T}(\Lambda)\right)\mathcal{T}^{\alpha\nu}(\Lambda) 
\nonumber\\&&-\frac{1}{\Lambda^4}\,\eta_{\beta\gamma}\mathcal{T}^{\alpha\beta}(\Lambda)
\partial_\alpha\mathcal{T}^{\gamma\nu}(\Lambda)
+ \mathcal{O}\left(\Lambda^{-6}\right).
\end{eqnarray}
This form is special in that it can be recognized as a covariant conservation law in the
effective, scale-dependent, metric $g_{\mu\nu}(\Lambda)$, defined as
\begin{equation}\label{gfic}
g_{\mu\nu}(\Lambda) = \eta_{\mu\nu}+ \frac{1}{\Lambda^4}\,\eta_{\mu\rho}\mathcal{T}^{\rho\sigma}(\Lambda)\eta_{\sigma\nu} +\mathcal{O}\left(\Lambda^{-6}\right).
\end{equation}
Indeed, \eqn{compspecial} can be rewritten as
\begin{equation}\label{magic}
\nabla_{(\Lambda)\mu} \mathcal{T}^{\mu\nu}(\Lambda) = 0,
\end{equation} 
where $\nabla_{(\Lambda)}$ is built from the metric \eqref{gfic}. Thus the equation for $\mathcal{T}^{\mu\nu}(\Lambda)$ takes exactly the same form as  the usual conservation equation $\partial_\mu\mathcal{T}^{\mu\nu}= 0$, but in a new, effective, metric.

Therefore the notion of ``highly efficient RG flow'' is that under this RG flow the equations obeyed by the 
projected operators take the same form as in the ultraviolet, but with an effective background metric which varies
with the RG scale in such a way to absorb the contributions from multi-trace operators to the respective equations.
Besides this effective metric, the equations for the projected operators can also include
 additional sources coupling to single-trace operators, which are scale-dependent as well\footnote{The idea that the sources coupling to single-trace operators become dynamical when the multi-trace operators are integrated out leading to classical gravity in one higher dimension has been explored before in the path-integral viewpoint before \cite{Heemskerk:2010hk,
Faulkner:2010jy,Lee:2013dln}. The approach described here takes the Heisenberg picture point of view.}. 

One can reconstruct the dual classical gravity theory from this ``highly efficient RG flow'' as follows. After taking the expectation value of  \eqn{gfic}  (assuming the large-$N$ factorization of multi-trace operators) and identifying the radial coordinate $z$ in the bulk spacetime with the scale $\Lambda$ via $z=\Lambda^{-1}$, the effective metric $g_{\mu\nu}(\Lambda)$ becomes essentially the induced metric on the hypersurface $z=\Lambda^{-1}$ in the bulk spacetime. Here we have assumed that the boundary (the UV of the dual theory) is at $z= 0$, so that the radial coordinate is the inverse of the scale. Nothing much changes if we do a radial reparametrization to put the boundary at $r=\infty$, using as for instance $r= l^2/z$, except that the scale $\Lambda$ now should be related to $r$ by $r=l^2\Lambda$. The effective equation \eqref{magic} is then equivalent to the so-called momentum constraints of the equations of classical gravity, evaluated on this hypersurface, with $\mathcal{T}^{\mu\nu}(\Lambda)$ being related to the extrinsic curvature of the hypersurface. 

In order to reconstruct the full bulk spacetime metric, one needs to decipher the automorphism group of the renormalization group flow which is related to the lift of the ultraviolet Weyl symmetry to an arbitrary scale. It can be shown that the latter completely encodes the information of the corresponding gauge fixing of the diffeomorphism symmetry in the bulk \cite{Behr:2015yna}. Using this information, one can recover the complete bulk spacetime metric from the renormalization group flow with an additional boundary condition given by the boundary metric, which corresponds to the actual metric on which the dual field theory lives \cite{Kuperstein:2011fn,Kuperstein:2013hqa}. This leads to the understanding that the holographic correspondence is equivalent to an existence of a special renormalization group flow in the field theory that is well defined by a set of principles.

The above discussion applies for purely holographic field theories. Our semi-holographic model can be motivated via three simple postulates:
\begin{itemize}
\item A highly efficient RG flow in the sense described here can be constructed only for a subset of soft degrees of freedom.
\item The effective background for these subset of degrees of freedom in the ultraviolet (i.e. initial conditions for the RG flow of this subset of degrees of freedom) are given by expectation values of operators projected to the hard sector.
\item An effective intermediate scale (like the saturation scale $Q_s$ here) controls the way the hard sector determines the effective background for the soft sector in the ultraviolet.
\end{itemize}
These postulates motivate our model, more particularly the action \eqref{ac} entirely. A more precise formulation is beyond the scope of this paper.

Let us finally notice that we have already used the above perspective in reinterpreting equations like \eqref{consv},
which were {\em a priori} obtained within holography and express conservation laws in a fictitious, curved,
space-time metric (similarly to \eqn{magic}), as ordinary equations of motion in flat space-time, 
but with additional sources describing the coupling to the hard sector, cf. \eqn{consv2}. As compared to \eqn{compspecial}, the right hand side of \eqn{consv2} does not
explicitly contain terms non-linear in the soft operator $\mathcal{T}^{\mu\nu}$, but only its couplings to hard sector observables, like $\st_{\mu\nu}$. Yet, such non-linear effects in $\mathcal{T}^{\mu\nu}$ are eventually generated via the backreaction of the soft sector on the hard fields, as encoded in the self-consistent solutions to the coupled equations
of motion for the two types of degrees of freedom.

\section{A semi-holographic model for jets}

In this section, we shall extend our semi-holographic model to also include the dynamics of 
a `jet' --- more precisely, a very heavy quark in the fundamental representation of the QCD gauge group SU$(N_c)$ --- propagating through the non--equilibrium quark--gluon plasma. The heavy quark can couple to both the soft
and the hard degrees of freedom of QCD, and in the present context these interactions should be modeled differently.
For the coupling to the soft modes which are strongly coupled, we shall use the standard holographic description
of a fundamental heavy quark as an open, semi-classical, Nambu-Goto string in AdS$_5$, with an endpoint attached
to the boundary\footnote{More precisely, the string endpoint is attached to a D7-brane which extends along the radial 
direction towards the interior of  AdS$_5$, from the boundary at $r\to\infty$ down to a value $r_M$
which is proportional to the bare mass $M$ of the quark. Here, however, we shall 
assume an infinitely heavy quark, for simplicity, hence $r_M\to\infty$ and the D7-brane plays no dynamical
role --- it can be identified with the  boundary of AdS$_5$.} 
(see e.g. Refs.~\cite{Herzog:2006gh,Gubser:2006bz,CasalderreySolana:2011us}).
For the interactions with the hard modes, which are weakly coupled, we shall use a semi-classical approximation,
in line with the fact that the hard modes themselves are described semi-classically, as
Yang-Mills fields on the boundary. Therefore, we shall treat the heavy quark partly also as a classical particle with a colour charge subjected to the non-Abelian Lorentz forces arising due to presence of the background Yang-Mills fields $A_\mu^a$ of the glasma, and with the colour charge additionally satisfying the Wong equation \cite{Wong}. To that aim, we shall assume that the
endpoint of the string at the AdS$_5$ boundary carries a non--Abelian color charge $\mathcal{Q}_a$,
which is a vector of SU$(N_c)$ with fixed normalization $ \mathcal{Q}_a \mathcal{Q}_a=C_F\equiv
(N_c^2-1)/2N_c$ and whose orientation can precess in time under the action of the color field
$A_\mu^a$ (see \eqn{jetac2} below). Note that this classical color charge is an additional ingredient
of the model, distinct from the conserved flavor charge of the heavy quark which will not be explicitly
considered in what follows, as it plays no special role under the assumed circumstances\footnote{The conservation
of the flavor charge reflects a world-volume gauge invariance
of the D7-brane, dual to the global flavor symmetry of the underlying gauge theory with fundamental
quarks. Here, however, we are not
interested in adding sources for the soft flavor current as the hard background flavour currents are assumed to be vanishing. Thus the bulk gauge fields dual to the soft flavor currents
also vanish. In the probe limit, the flavour charge at the end-point of the quark thus play no role in the dynamics in the absence of background hard and soft flavour currents.}. However it can be
easily added to the discussion if one is interested.

The string embedding in the bulk spacetime is given by $X^M(\tau, \sigma)$, where 
$X^M =(r, x^\mu)$ are the coordinates of the bulk spacetime, whereas
$\sigma^\alpha = (\tau, \sigma)$ are the string world-sheet coordinates.
The world-sheet time $\tau$ parametrizes the boundaries which are world-lines. The boundary $\sigma=0$, in particular, will correspond to the end point of the string fixed on the boundary of $AdS_5$, which is given by $r\to\infty$ 
in the Eddington--Finkelstein coordinates that we shall use throughout this section.

The semi-holographic model for the dynamics of the open string is encoded in the following action:
\begin{eqnarray}\label{jetac1}
S_{\rm string} &=& -T_0\int {\rm d}^2 \sigma  \, \sqrt{-{\rm det} \, h_{\alpha\beta}} +  g\int_{\sigma = 0} {\rm d}\tau \, \mathcal{Q}_a (\tau) \, \, \frac{ {\rm d}x^\mu(\tau)}{ {\rm d}\tau} \, \, A^{a}_{\mu}(X(\tau)),
\end{eqnarray}
where $T_0$ is the string tension and $h_{\alpha\beta}$ is the induced world--sheet metric, which reads
\begin{equation}
h_{ab} = G^{{\rm (s)}}_{MN}\partial_\alpha X^M \partial_\beta X^N,
\end{equation}
with the string--frame metric $G^{{\rm (s)}}_{MN}$ related to the bulk--metric $G_{MN}$ and the bulk dilaton $\Phi$ via
\begin{equation}\label{stringmetric}
G^{{\rm (s)}}_{MN} = e^{\frac{\Phi}{2}}G_{MN}.
\end{equation}
It is not hard to see that $T_0$ enters the equation of motions only in the combination $T_0 l^2$, where $l$ is the radius of the ambient anti-de Sitter space. Therefore, we have one additional dimensionless parameter, namely $T_0l^2$, for describing jet dynamics.

In order for the action to be gauge--invariant, the non--Abelian charge $\mathcal{Q}_a$ at the end of the string should follow the following condition (below, $x^\mu(\tau) \equiv x^\mu(\tau, \sigma=0)$)
\begin{equation}\label{jetac2}
\frac{ {\rm d}\mathcal{Q}_a(\tau)}{ {\rm d}\tau} +g f_{abc} \frac{ {\rm d}x^\mu(\tau)}{ {\rm d}\tau}A_\mu^b(x(\tau))\mathcal{Q}^c(\tau) = 0,
\end{equation}
which is recognized as the Wong equation for the precession of the color charge
\cite{Wong,Litim:2001db}.

For the physical interpretation and also for practical calculations, 
it is useful to chose the world--sheet coordinates as
\begin{equation}\label{gaugefix}
\tau = t, \quad \sigma = r,
\end{equation}
where $t$ and $r$ are the time and radial coordinates of the embedding spacetime. In this case, the bulk coordinates $X^M(\tau,\sigma)$ become $\left(t, r, x^i(t,r)\right)$. We recall that the radial coordinate $r$ extends from $\infty$ up to the apparent horizon at 
$l^2\Lambda_{{\rm QCD}}$. In this gauge, the world--sheet action becomes
\begin{eqnarray}\label{jetac1gf}
S_{\rm string} &=& T_0\int {\rm d} t\int_{l^2\Lambda_{{\rm QCD}}}^\infty {\rm d}r \, \sqrt{-{\rm det} \, h_{\alpha\beta}} +  g\int_{r = \infty} {\rm d}t \, \mathcal{Q}_a (t) \, \, \frac{ {\rm d}x^\mu(t)}{ {\rm d}t} \, \, A^{a}_{\mu}(x(t))\, .
\end{eqnarray}
The last piece in the action is then recognized as the standard interaction between a (classical)
color current and the non--Abelian gauge field, that is,  $g\int \rmd^4x \,j^\mu_a(x) A^{a}_{\mu}(x)$ with
\beq\label{current}
 j^\mu_a(x)\,\equiv \,g \int {\rm d}t \,  \, \mathcal{Q}_a (t) \, \, \frac{ {\rm d}x^\mu(t)}{ {\rm d}t} \, \, 
 \delta^{(4)}\big(x-x(t)\big)
 \, .
 \eeq
Furthermore, the Wong equation  \eqref{jetac2} can be understood as the covariant conservation law for the current
in the presence of the classical Yang--Mills field: $D_t Q=0$, where $D_t=\frac{ {\rm d}x^\mu(t)}{ {\rm d}t}
D_\mu$ is the covariant derivative along the quark world line and $D_\mu^{ab}[A]=\del_\mu\delta^{ab}+
gf^{abc}A_\mu^c$ the covariant derivative in the adjoint representation. This equation preserves the norm of the
adjoint color vector $ \mathcal{Q}_a$, as it should.

After adding the string action 
\eqref{jetac1gf} to the glasma action \eqref{ac}, it is clear that the color current $ j^\mu_a(x)$ of the heavy quark
counts as an additional source in the r.h.s. of the Yang--Mills equations \eqref{glback}. By assumption, the heavy quark represents only a small perturbation of the medium, so the change in the glasma field 
induced by this current can be computed in the linear response approximation. The small
correction $\delta A^\mu$ obtained in this way, which is linear in $ j^\mu$, is important for our present
purposes, in that it encodes a part of the energy loss by the heavy quark --- namely, that part
due to radiation which is induced by its interactions with the hard, glasma fields. In a purely perturbative
set--up at weak coupling, a similar calculation provides the medium induced radiation {\em a la} BDMPSZ
\cite{Baier:1996kr,Baier:1996sk,Zakharov:1997uu}. Note that the interactions relevant for
this mechanism involve both the color precession of the color current  $ j^\mu_a(x)$ 
of the heavy quark, as described by  \eqn{jetac2}, and the non--linear effects in the Yang--Mills equations 
\eqref{glback}, which describe the scattering between the radiated gluon 
(the induced field $\delta A^\mu$) and the hard modes in the glasma (the background
field $A^\mu$).

In the present set--up, there are two additional mechanisms leading to energy loss by the heavy quark.
First, there is collisional energy loss associated with the scattering off the hard partons; in the present
approximations, this is described as the acceleration of the heavy quark by the Lorentz force generated by
the glasma fields. Second, there is another type of medium--induced radiation, which is generated by the
interactions between the heavy quark and the soft, strongly--coupled, modes; in our model, 
this is holographically represented as the energy flow down the string world--sheet. To 
compute these various effects, one needs the trajectory $x^i(t)$ of the heavy
quark (the string endpoint at the Minkowski boundary) and, more generally, the string world--sheet 
$X^M(\tau,\sigma)$ in the bulk.
For that purpose, we need to solve the Nambu--Goto equations with appropriate boundary conditions. 

To construct these equations and the associated boundary conditions, 
it is preferable to work with manifestly geometric notations,
which are reparametrization--invariant. Hence, we return to our generic world--sheet
coordinates $\tau$ and $\sigma$, with the latter extending from $\sigma=0$ (the endpoint of the string at $r=\infty$)
to $\sigma=\sigma_{{\rm max}}$ (the endpoint of the string at $r=l^2\Lambda_{{\rm QCD}}$). 
The variation of the action gives 
\begin{eqnarray}\label{variation}
\delta S_{\rm string} &=& -\int {\rm d} \tau \int_0^{\sigma_{{\rm max}}} {\rm d} \sigma \, \sqrt{-{\rm det} \,\,h} \,\, \mathcal{D}_\alpha P^{\alpha}_{\,\, M}\delta X^M\nonumber\\ && +\int_{\sigma=0} {\rm d} \tau  \left( \sqrt{-h_{\tau\tau}} \, n_\alpha P^\alpha_{\,\,M}\delta X^M+  g\mathcal{Q}_a \frac{{\rm d}x^\nu}{{\rm d}\tau}F^a_{\mu\nu}\delta x^\nu\right)
\nonumber\\ && +\int_{\sigma=\sigma_{{\rm max}}} {\rm d} \tau \left(\sqrt{-h_{\tau\tau}}\,n_\alpha P^\alpha_{\,\,M}\delta X^M\right).
\end{eqnarray}
We have used the Wong equation \eqref{jetac2} in order to obtain the above form. Also
\begin{equation}
P^\alpha_{\,\,M} = -T_0  G^{{\rm (s)}}_{MN}h^{\alpha\beta}\partial_\beta X^N
\end{equation}
is the spacetime energy--momentum current carried by the string, $\mathcal{D}$ is the world--sheet covariant derivative built from the induced world--sheet metric $h$. Furthermore, $n_\alpha$ is the normal to the world--sheet boundary; explicitly it reads
\begin{equation}
n_\alpha =\left(0, \frac{\sqrt{-{\rm det}\,h}}{\sqrt{-h_{\tau\tau}}}\right). 
\end{equation} 
In particular, in the gauge \eqref{gaugefix}, this becomes
\begin{equation}
\sqrt{-h_{\tau\tau}}\,n_\alpha P^\alpha_{\,\,M} = \sqrt{-{\rm det} \, h}\,P^r_{\,\,M}.
\end{equation}

The string equation of motion which follows from the first line of \eqref{variation} is
\begin{equation}\label{eomstring}
\mathcal{D}_\alpha P^\alpha_{\,\, M} = 0.
\end{equation}

The end of the string at $\sigma=0$ is fixed at the Minkowski boundary $r=\infty$. Therefore, 
in a generic gauge, the coordinate $r(\tau, \sigma)$ of the string satisfies the Dirichlet boundary condition
\begin{equation}\label{bc1sigma=0}
r(\tau, \sigma = 0)  =\infty
\end{equation}
for all times $\tau$. Notice that this boundary condition is indeed compatible with the second line of \eqref{variation} giving the variation of the world--sheet action. In the particular gauge \eqref{gaugefix}, $r$ is no longer a dynamical variable, as it becomes the world--sheet spatial coordinate. In that gauge, \eqn{bc1sigma=0}
is not a boundary condition anymore, but it merely specifies one of the string endpoints.

Still at $\sigma=0$, the other  coordinates $x^\mu$ satisfy the boundary condition
\begin{equation}\label{bc2sigma=0}
\sqrt{-h_{\tau\tau}} \, n_\alpha P^\alpha_{\,\, \mu}+  g\mathcal{Q}_a \frac{{\rm d}x^\nu}{{\rm d}\tau}F^a_{\mu\nu}= 0.
\end{equation}
Once again, this is compatible with the second line of \eqref{variation}. Physically, \eqn{bc2sigma=0}
expresses the generalization of the equation of motion of a charged particle under the influence of the non--Abelian Lorentz force generated by the glasma fields. In our set--up, this force acts at the endpoint of the string and injects energy and momentum into the string in the Minkowski directions. Notice also that the string endpoint lives in a non--trivial space--time metric, cf. \eqn{sc}, due to the boundary sources representing hard fields\footnote{Note that $G^{{\rm(s)}}_{\mu\nu}$ blows up at the boundary $r=\infty$ like $r^2$, however $P^r_{\,\,\mu}$ remains finite because $\partial_\sigma x^i(\tau, r)$ (or $\partial_r x^i(\tau, r)$ in the gauge \eqref{gaugefix}) vanishes at the boundary with the right power of $r$. This means that the string always ends perpendicularly at the boundary. This is a 
familiar feature for trailing string solutions \cite{Herzog:2006gh,Gubser:2006bz}.}.


We furthermore need to set boundary conditions at $\sigma = \sigma_{{\rm max}}$. For the radial coordinate $r$,
we should clearly require
\begin{equation}\label{bcsigma=pi}
r(\sigma = \sigma_{{\rm max}}, \tau) = l^2\Lambda_{{\rm QCD}}\,,
\end{equation}
which is indeed consistent with the variation of the action (cf. the third line in \eqref{variation}).
(In the gauge \eqref{gaugefix},  this boundary condition on $r$ should be thought of as a mere 
specification of the respective endpoint.)
Regarding the other coordinates $x^\mu$, it is useful to recall the finding in previous studies
\cite{Herzog:2006gh,Gubser:2006bz,Gubser:2006nz,CasalderreySolana:2007qw,Giecold:2009cg}, 
that the respective boundary conditions should not be set
at the apparent horizon (which in our notations correspond to $\sigma = \sigma_{{\rm max}}$), but 
rather at the {\em world--sheet horizon}. As known, a 
world--sheet horizon is bound to  form whenever there is acceleration and/or an event 
horizon in the embedding space--time \cite{Chernicoff:2008sa,Dominguez:2008vd,Xiao:2008nr}. 
In our case, both situations may occur in general, 
as an event horizon dynamically emerges in $AdS$ 
and moreover the endpoint of the string on the boundary can
be accelerated by the non--Abelian Lorentz force. In general the world--sheet horizon is 
located {\em above}  the event horizon (i.e. closer to the boundary at $r=\infty$), but it may 
also coincide with the latter in special situations (e.g. for a string at rest). The emergence 
of a world--sheet horizon manifests itself as a singularity in the solution to
the Nambu-Goto equation. Whenever this happens, one must choose the
integration constant to ensure that this horizon is a regular point on the string, and 
not an endpoint (see \cite{Herzog:2006gh,Gubser:2006bz} for an explicit realization of this idea
in the case of the trailing string).  Notice that the piece of the string which is located below the
world--sheet horizon is causally disconnected from the string endpoint at the boundary. Accordingly,
the energy and momentum fluxes across this world--sheet horizon should be interpreted as
the energy and momentum losses by the heavy quark via radiation of soft and strongly--coupled quanta  
\cite{Dominguez:2008vd,Xiao:2008nr,Giecold:2009cg,Hatta:2008tx,Hatta:2011gh,Garcia:2012gw}.

To fully specify the solution to the Nambu-Goto equation  \eqref{eomstring}, we also need initial conditions,
say, at $t=t_0$. For all
 the string points except for its endpoint at $\sigma = 0$, this initial condition refers to
both the position and the velocity of the string points in the bulk. At $\sigma = 0$, 
however, it is sufficient to specify the position $x^i(t_0)$ of that endpoint: indeed, the boundary condition \eqref{bc2sigma=0} determines the respective velocity including at the initial time, 
provided the initial condition for the color 
vector $\mathcal{Q}_a$ is also specified. 

In this respect, our set--up differs from other approaches, which employ the Dirichlet boundary condition  
at $\sigma = 0$, meaning that one needs to specify the entire trajectory $x^i(t)$
of the string endpoint on the boundary of $AdS_5$, at any $t$. For instance, in Refs.~\cite{Herzog:2006gh,Gubser:2006bz}, the string 
endpoint on the boundary is assumed to move at a constant velocity, under the combined action of its 
interactions with  the plasma and of an external force.  In that case, $P^r_{\,\,\mu}$ is calculated from the string profile, to obtain the (velocity--dependent) force acting on the string, which is then interpreted as (minus)
the drag force acting on the heavy quark from the medium. In our approach, the endpoint of the string is 
fully dynamical and the energy loss (in particular the stopping distance) is obtained by solving
the equations of motion.
 
In our particular physics set--up, where the soft sector is dynamically created at $t>0$ by the hard sources,
it is convenient to chose the initial conditions for the string at the collision time $t=0$. A simple choice
is to assume that, at $t=0$, we have a vertical string extending from $r=\infty$ down to 
$r= l^2\Lambda_{{\rm QCD}}$, and moving with a given velocity which is uniform along the string. 
This corresponds to an heavy quark with a specific initial kinetic energy 
in the center--of--mass frame of the collision.

To summarize, the string equations of motion \eqref{eomstring} and the Wong equation \eqn{jetac2} for the non-Abelian charge attached to the end-point at the boundary of $AdS$, supplemented with appropriate
initial conditions, the boundary conditions \eqref{bc1sigma=0}, \eqref{bc2sigma=0},
and with the regularity condition at the world--sheet horizon as discussed above, should uniquely specify the 
dynamics of the string -- particularly the energy--momentum loss towards the soft,
strongly--coupled, sector. The string solution will also provide the trajectory $x^\mu(t)$ of the heavy quark (meaning the trajectory of the end-point at the boundary of $AdS$) determining the time evolution of the non-Abelian charge current \eqref{current}, which then can be used to calculate the additional energy loss towards the hard modes.
In the future, we would like to develop an explicit algorithm for  numerically solving the  dynamics
of this semi--holographic jet. The equivalent Polyakov action formulation as employed in
\cite{Chesler:2013cqa} might be helpful in this regard.

\section{Perspectives and future directions}

In principle, our semi-holographic model can be used to calculate a wealth of experimental observables. In general, though, the associate numerical simulations may turn out to be rather cumbersome in practice. It would be therefore useful to dispose of a simplified picture, particularly for the late stages of the dynamics, where a state
of local thermal equilibrium has been presumably reached and a hydrodynamical effective description becomes
appropriate. It is indeed well known that, as far as the holographic part is concerned, 
the late time dynamics can be captured by an asymptotic hydrodynamic expansion with arbitrarily large number of transport coefficients \cite{Janik:2005zt,Baier:2007ix,Bhattacharyya:2008jc,Heller:2013fn}. On the other hand the classical Yang-Mills equations are known to reduce to transport equations of the Boltzmann type
as soon as the occupation numbers for the hard gluon modes become much smaller than $1/\alpha_s$
(which eventually occurs due to the longitudinal expansion) 
\cite{Mueller:2002gd,Mathieu:2014aba,Epelbaum:2014mfa}. 
By appropriately combining these observations, one should be able to simplify our semi-holographic model in the final stages of the expansion, before the onset of hadronization.

The Boltzmann equation for the hard gluons emerging from our generalized Yang--Mills equations should
be more general than the standard respective equation, in the sense of including additional contributions
to the collision term which are associated with the exchange of bulk hydrodynamic modes via hard--soft couplings. Similarly, the hydrodynamics of the soft sector should be considered in a different effective metric that is modified as in \eqref{bcs}, where the $t_{\mu\nu}$ of the hard-sector will be obtained from the Boltzmann equation. Thus we will obtain a novel and relatively simple description involving Boltzmann equations for the gluons coupled with the hydrodynamics of the fluid composed of the soft degrees of freedom, and living in an effective metric that is in turn determined by the gluon distribution which solves the Boltzmann equation. The only input of holography will be to provide the transport coefficients of the fluid composed of the soft degrees of freedom. Thus we reduce the problem of solving classical gravity equations to that of solving hydrodynamics living in a non-trivial background. At the same time,
the classical Yang-Mills equations are replaced by a Boltzmann equation with a modified collision term. We hope that such a simplified picture will be sufficient to study observables like the collective flow.

Recently it has been shown that a generalization of the thermal fluctuation-dissipation relation \cite{Banerjee:2012uq,Mukhopadhyay:2012hv} also holds for {\em non-equilibrium} holographic states which relax to static 
equilibrium and are sufficiently close to it --- in the sense that one can describe the state by using either a
hydrodynamic expansion in powers of derivatives, or an expansion in powers of the amplitude
of the perturbation  \cite{Iyer:2009in,Iyer:2011qc,Heller:2013oxa}. Some recent results indicate that such 
a generalized fluctuation-dissipation relation may also hold for the holographic system approaching a boost-invariant perfect fluid expansion and can be used to compute e.g. prompt photon and dilepton productions from the non--equilibrium QGP \cite{Baier:2012ax,Baier:2012tc,Mamo:2014ema}.
We hope to be able to design more direct tests of this state-independent non-equilibrium fluctuation-dissipation relation for the holographic sector at late time, by correlating suitable observables. 
 
To conclude, more work is needed both on the theory side, to construct self--consistent numerical solutions to the coupled hard--soft equations and develop  
simplified versions of the model which are suitable at late time;
and on the phenomenology side, to identify and compute new observables which are relevant for the physics
of heavy-ion collisions. We intend to address such tasks in future work.

\acknowledgments

  We are grateful to Al Mueller for insightful discussions. We thank Florian Preis and Anton Rebhan for many helpful comments on the manuscript. The research of E.I. is supported by the European Research Council under the Advanced Investigator Grant ERC-AD-267258. The research of A.M. involving this work has been supported by the LABEX P2IO, the ANR contract 05-BLAN- NT09-573739, the ERC Advanced Grant 226371 and the ITN programme PITN-GA-2009-237920. Presently the research of A.M. is supported in part by European Union's Seventh Framework Programme under grant
agreements (FP7-REGPOT-2012-2013-1) no 316165, the EU-Greece
program "Thales" MIS 375734 and was also co-financed by the European
Union (European Social Fund, ESF) and Greek national funds through the Operational
Program "Education and Lifelong Learning" of the National Strategic Reference
Framework (NSRF) under "Funding of proposals that have received a positive
evaluation in the 3rd and 4th Call of ERC Grant Schemes".


\bibliographystyle{utcaps}
\bibliography{semihol-refs}



\end{document}